\documentclass[%
 aip,
 amsmath,amssymb,
 reprint,%
]{revtex4-2}

\usepackage{graphicx}
\usepackage{dcolumn}
\usepackage{bm}
\usepackage{float}

\usepackage[utf8]{inputenc}
\usepackage[T1]{fontenc}
\usepackage{mathptmx}
\usepackage{etoolbox}
\usepackage{amsmath, amssymb}

\makeatletter
\def\@email#1#2{%
 \endgroup
 \patchcmd{\titleblock@produce}
  {\frontmatter@RRAPformat}
  {\frontmatter@RRAPformat{\produce@RRAP{*#1\href{mailto:#2}{#2}}}\frontmatter@RRAPformat}
  {}{}
}%
\makeatother
\begin{document}

\preprint{AIP/123-QED} 

\title[NRSWB]{High wave vector non-reciprocal spin wave beams} %

\author{L. Temdie}%
 \altaffiliation[Also at ]{Lab-STICC - UMR 6285 CNRS, Technopole Brest-Iroise CS83818, 29238 Brest Cedex 03 France}
 \author{V. Castel}%
 \altaffiliation[Also at ]{Lab-STICC - UMR 6285 CNRS, Technopole Brest-Iroise CS83818, 29238 Brest Cedex 03 France}

\affiliation{ 
IMT- Atlantique, Dpt. MO, Technopole Brest-Iroise CS83818, 29238 Brest Cedex 03 France
}%
\author{C. Dubs}%
\affiliation{ 
INNOVENT e.V. Technologieentwicklung, Pruessingstrasse 27B, 07745 Jena, Germany}

\author{G. Pradhan}
\author{J. Solano}
\author{H. Majjad}
\author{R. Bernard}
\author{Y. Henry}
\author{M. Bailleul}
\affiliation{%
IPCMS - UMR 7504 CNRS
Institut de Physique et Chimie des Matériaux de Strasbourg France
}%

\author{V. Vlaminck}
 \email{vincent.vlaminck@imt-atlantique.fr}
 \altaffiliation[Also at ]{Lab-STICC - UMR 6285 CNRS, Technopole Brest-Iroise CS83818, 29238 Brest Cedex 03 France}

\affiliation{ 
IMT- Atlantique, Dpt. MO, Technopole Brest-Iroise CS83818, 29238 Brest Cedex 03 France
}%

\date{\today}

\begin{abstract}
We report unidirectional transmission of micron-wide spin waves beams in a 55 nm thin YIG. We downscaled a chiral coupling technique implementing Ni$_{80}$Fe$_{20}$ nanowires arrays with different widths and lattice spacing to study the non-reciprocal transmission of exchange spin waves down to $\lambda\approx$ 80\,nm. A full spin wave spectroscopy analysis of these high wavevector coupled-modes shows some difficulties to characterize their propagation properties, due to both the non-monotonous field dependence of the coupling efficiency, and also the inhomogeneous stray field from the nanowires.   
\end{abstract}

\maketitle


\section{INTRODUCTION} 

Non-reciprocity is an essential properties of today's information systems \cite{Palmer2019}. The ability to inhibit signal flow in one direction while allowing it in the reverse direction is crucial to either protect devices from reflection, isolate transmitters and receivers in radar architecture, or even shield qubits from its environment in quantum computers. Most of the current non-reciprocal functionalities rely on the gyrotropic nature of the magnetization dynamics in field-based ferrimagnetic systems (namely Yttrium Iron Garnet - YIG), which tend to be large, and costly to assemble \cite{Harris2012}. Future progress in communication systems, and most critically in quantum technologies, rely heavily on the possibility to miniaturize and integrate these non-reciprocal devices \cite{Devoret2013}.\\
The field of magnonics, which implements magnetic excitation called spin waves -or their quanta magnons-, is actively involved in the search of non-reciprocal scalable solutions\cite{Chumak2022}. Extensive research in the last decade revealed the possibility to engineer both amplitude and frequency non-reciprocity of spin waves in many different ways \cite{Barman2021}. Firstly, the well-known Damon-Eshbach (DE) configuration, where the equilibrium magnetization of a thin film lies in the plane of the film and perpendicular to the wavevector \cite{DE}, displays non-reciprocal dynamic amplitude across the thickness for oppositely traveling waves, which couple chirally to an excitation antenna \cite{Camley1987}. Moreover, this non-reciprocity was recently proven to be strongly enhanced in the presence of magnetic nanostructures, where unidirectional transmission of spin waves was achieved using the chiral coupling between the FMR of Co nanowires and exchange spin waves in a thin YIG film \cite{Chen2019,Wang2021}. Secondly, small frequency non-reciprocity (namely f(k)\,$\ne$\,f(-k)) was demonstrated more recently in various systems, either with asymmetrical surface anisotropies between top and bottom surfaces \cite{Gladii2016}, or with the coupled dynamics of ferromagnetic multilayers \cite{Grassi2020}, or also with the interfacial Dzyaloshinskii-Moriya interaction (DMI) of a ferromagnetic layer coupled to a high spin-orbit material \cite{Moon2013,Brächer2017,Di2015}. Additionally, asymmetric spin-wave dispersion was recently predicted in non-planar geometries due to a topographically induced dynamic dipolar effect \cite{Otalora2016}. Nevertheless, all these frequency non-reciprocity effects only becomes significant when the magnons wavelength reaches sub micrometric sizes. Lastly, non-reciprocal functionalities have also been predicted very recently using an innovative inverse design approach \cite{Qi2021,Papp2021}. \\
In this communication, we further miniaturized the method of Chen et al. \cite{Chen2019} and demonstrate the possibility of shaping non-reciprocal spin wave beams in a continuous thin YIG film. 
This paper is organized as follows: in Sec. II, we present the sample design and the experimental protocol used to measure broadband multi-modes spin waves transmission. In Sec. III, we present the non-reciprocal spin wave beams spectra, and address through a spin wave spectroscopy study the peculiarities encountered in shaping narrow spin wave beams via this method.  \\

\begin{figure}
\centering
\includegraphics[width=80mm]{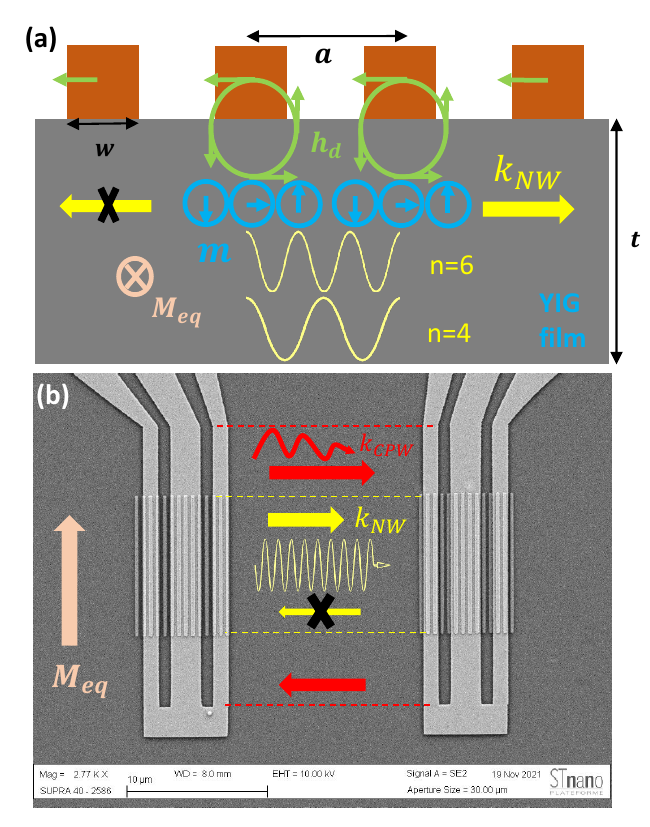}
\caption{\label{fig1} (a) Sketch of the NWA-FMR mediated chiral coupling mechanism. (b) SEM image of device\,I,  $w$=300\,nm width and $a$=500\,nm lattice constant of  Ni$_{80}$Fe$_{20}$ NWA grown on top of 55 nm thickYIG film, 80 nm Au-antennas grown on top of YIG|Ni$_{80}$Fe$_{20}$. Red and Yellow color represent respectively antenna ($k_{CPW}$) and  Ni$_{80}$Fe$_{20}$ NWA($k_{NW}$) excite mode .}
\end{figure}

\begin{figure*}[t]
\includegraphics[width=160mm]{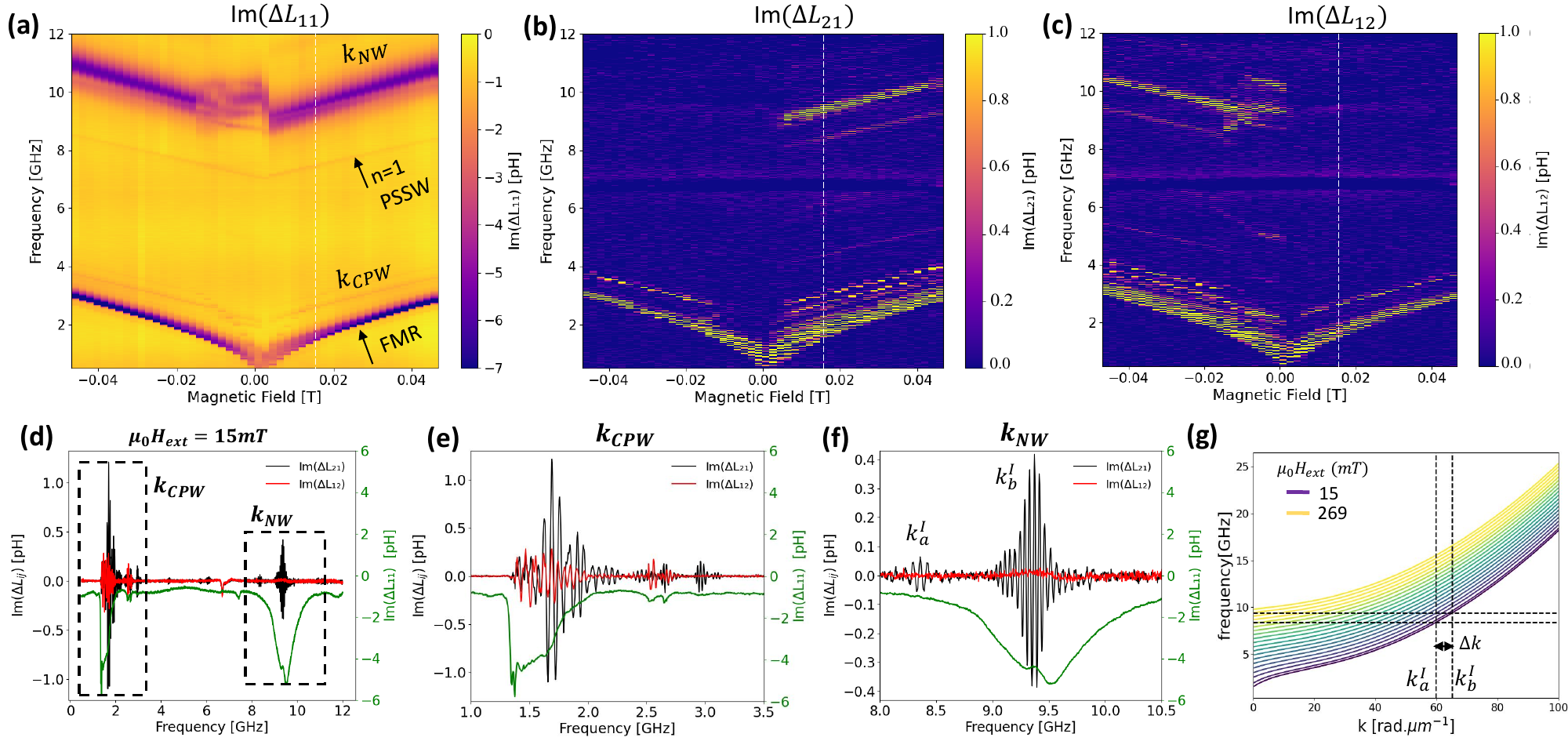}
\caption{\label{fig2} (f,H) mapping of the spin wave spectra obtained for device\,I for (a) $\Delta L_{11}$, (b) $\Delta L_{21}$, and (c) $\Delta L_{12}$. (d) Spectra obtained at $\mu_0$H$_{ext}$=+15\,mT from (b) and (d). (e) Zoom of (d) on the $k_{CPW}$ region, and (f) on the $k_{NW}$ region. (g) Dispersion relation for field ranging from 50\,mT to 269\,mT.}
\end{figure*}

\section{Experimental Set-Up}

\subsection{Sample design and fabrication}
The chiral excitation of propagating spin waves is achieved by coupling the ferromagnetic resonnance (FMR) of magnetic nanowires array (NWA) to a low damping continuous film such as YIG. As illustrated in Fig.1-(a), the phase profile of a propagating spin wave excited by the dynamic dipolar field of nanowires, all precessing in phase, only matches for a single propagation direction. The degree of chirality depends strongly on the ellipticity of the dynamic dipolar field of the NWA \cite{Yu2021}. Typically, the elliptical polarization of the Kittel mode in flat rectangular nanowires breaks the perfect chirality. For this reason we made the nanowires 60\,nm thick to come as close as possible to an aspect ratio $t/w=1$ which produces circularly polarized dipolar field. Moreover, the Kittel mode frequency of the nanowires can be tuned up with decreasing their width $w$, therefore the NWA can excite propagating modes ($k_{NW}$) with much higher wavevector than the one directly coupled by the microwave field of the antenna ($k_{CPW}$). In the case of an extended array of NWA, these high vector ($k_{NW}$) correspond simply to integer values of $\frac{2\pi}{a}$ accordingly with the periodic boundary conditions of the phase precession \cite{Liu2018,Chen2018}, minus some possible extinction related to the ratio of $w/a$. In order to study the dependence of the NWA array density on the excitation efficiency of the spin wave beam, we designed two devices with different width $w$ and lattice constant $a$. Namely, for device\,I, $w$=300\,nm and $a$=500\,nm; and for device\,II, $w$=200\,nm and $a$=400\,nm. The NWA length of 10\,$\mu$m was kept the same for both devices. Such a localized distribution of excitation field produces a focused emission of spin waves in a similar fashion to the spin wave beam excited from a constricted CPW  \cite{Loayza2018,Körner2017}. In the present geometry, the distance of propagation is well within the near-field region defined as the ratio $L^2/\lambda_{NW}$, with $L$ being the antenna length and $\lambda_{NW}$ the magnon wavelength, so that the near-field diffraction pattern from such a localized excitation follows very closely its shape. For this reason, the emission of $k_{NW}$ magnons remains confined within the length of the NW, thus forming a spin wave beam of width closely equal to 10 µm as sketched in Fig. 1-(b). \\
Fig.1-(b) shows a SEM image of device\,I. The sample fabrication required two steps of e-beam lithography followed by e-beam evaporation lift-off process on a 55\,nm thick liquid phase epitaxy YIG film \cite{Dubs2020}. To circumvent the insulating nature of the substrate, we resorted to an extra layer of conductive resist AR Electra 92 \cite{Electra92} on top of PMMA. In the first step, we structured the 60\,nm thick Ni$_{80}$Fe$_{20}$ (Permalloy) using a 3\,nm thin Ti adhesion layer, and a 8\,nm Au capping layer via e-beam evaporation. In the second step, we aligned for both devices the same 20\,$\mu$m long coplanar wave guides (CPW) with the following lateral dimensions: 2\,$\mu$m wide central line and 1\,$\mu$m wide ground lines each spaced by 1\,$\mu$m with the central line. These pairs of spin wave antenna made of 4\,nm Ti + 80\,nm Au also via ebeam evaporation. The center-to-center distance D between the two CPW are respectively 20\,$\mu$m for device\,I, and 15\,$\mu$m for device\,II.

\subsection{Broadband Spin Wave Spectroscopy}
  
We incorporated an home-made confined electromagnet onto a PM8 probe station to perform VNA spin wave spectroscopy using a Rohde \& Schwarz ZNA43GHz Vector Network Analyzer, calibrated via a SOLT procedure with 150\,$\mu$m pitch picoprobes. The sample is carefully placed within the 11\,mm gap of the electromagnet, whose poles are 15\,mm in diameter, and produce an homogeneous in-plane field along the poles axis (namely less than 0.3\% variation at most) up to 500\,mT at 3\,A. Due to the small hysteresis of the electromagnet, we always initialize a measurement with a high current of 5A in order to be consistent with the calibration of our magnet.\\
We performed broadband frequency sweep in the [0.5;20]\,GHz range at constant applied field acquiring 3201 points with a resolution bandwidth of 100\,Hz in order to resolve  widely spread-out multi-modes transmission spectra. We measure 2-ports S-parameters for several field values, which we translate into the corresponding $Z_{ij}$ impedance matrices. Furthermore, due to the small area of NWA (namely 10*5$\mu$m$^2$), the typical range of $S_{ij}$ signal amplitudes are of the order of 10$^{-4}$ in linear scale (or -80\,dB in log scale), while the noise floor lays at 10$^{-6}$. We retrieve a flat base line by subtracting the measurement at given field with a reference measurement at another field value $H_{ref}$ far enough that there are no dynamic feature in the frequency span. Finally, as the coupling of the spin wave to an antenna is of inductive nature, we chose to represent our relative measurements in units of inductance defined as \cite{Vlaminck2010,Gladii2016_APL,Loayza2018}:     
\begin{equation}
\Delta L_{ij}(f,H)=\frac{1}{i\,2\pi\,f}(Z_{ij}(f,H)-Z_{ij}(f,H_{ref}))
\label{DeltaL}
\end{equation}
where the subscripts $(i,j)$=1\,or\,2 denote either a transmission measurement between two antennas ($i \ne j$), or a reflection measurement done on the same port ($i=j$).

\section{Experimental Results and discussion}
\subsection{Perfect non-reciprocal transmission}
We present in Fig.2 a mapping in the (f,H) plane of the spectra $\Delta L_{11}$, $\Delta L_{21}$, and $\Delta L_{12}$ acquired with device\,I for field gradually changing from +46\,mT to -46\,mT. We observe a myriad of modes that we can separated into two main regions. The first region at lower frequencies corresponds to all the $k_{CPW}$ modes coupled directly by the microwave field of the antenna, and which range typically from k$\approx$0 (namely the section of the CPW around the picroprobes) to k$\approx$15\,rad.$\mu$m$^{-1}$ (the 8$^{th}$ satellites peak of the antenna). As shown in Fig.2-(e) with the spectra measured at $\mu_0$H$_{ext}$=+15\,mT, some partial non-reciprocity occurs between $\Delta L_{21}$ and $\Delta L_{12}$ due to the ellipticity of the microwave field of the antenna as mentioned above. \\ 
The second region above 8 GHz corresponds to the higher $k_{NW}$ modes mediated by the FMR of the permalloy NWA. In this frequency region, Fig.2-(b) and 2-(c) show a perfect 100\% non-reciprocal transmission of spin waves. Namely, at positive field values, $\Delta L_{21}$ shows large oscillations, while no transmission occurs from port\,2 to port\,1. Conversely at negative fields, the non-reciprocity is reversed, and no transmission occurs from port\,1 to port\,2. We emphasize that the dynamic range employed for our measurements (iBW=100Hz, P=-10dBm) gives a noise floor of about 5\,fH, while typical transmission amplitude are of the order of 100\,fH. Besides, one notices between 0 and -15\,mT that the transmission for $\Delta L_{12}$ is somewhat dispersed, indicating that the magnetization of the permalloy NWA do not conserve an anti-parallel orientation with the YIG magnetization, and that a gradual switching of the Py NWA is likely occurring. Furthermore, as shown with the spectra at +15\,mT of Fig.2-(f), a closer look in this region shows an additional perfectly non-reciprocal mode around 8.4\,GHz on top of the more pronounced mode at 9.4\,GHz. Nevertheless, the frequency spacing between these two modes is much smaller than expected, as the wavevector difference between two adjacent modes, $k_{n+1}-k_{n}=\dfrac{2\pi}{a}$ would result in a frequency difference of about 3\,GHz, as can be deduced from Fig.2-(g). We investigate further the nature of these $k_{NW}$ multimodes in the next section. 

\begin{figure*}[t]
\includegraphics[width=160mm]{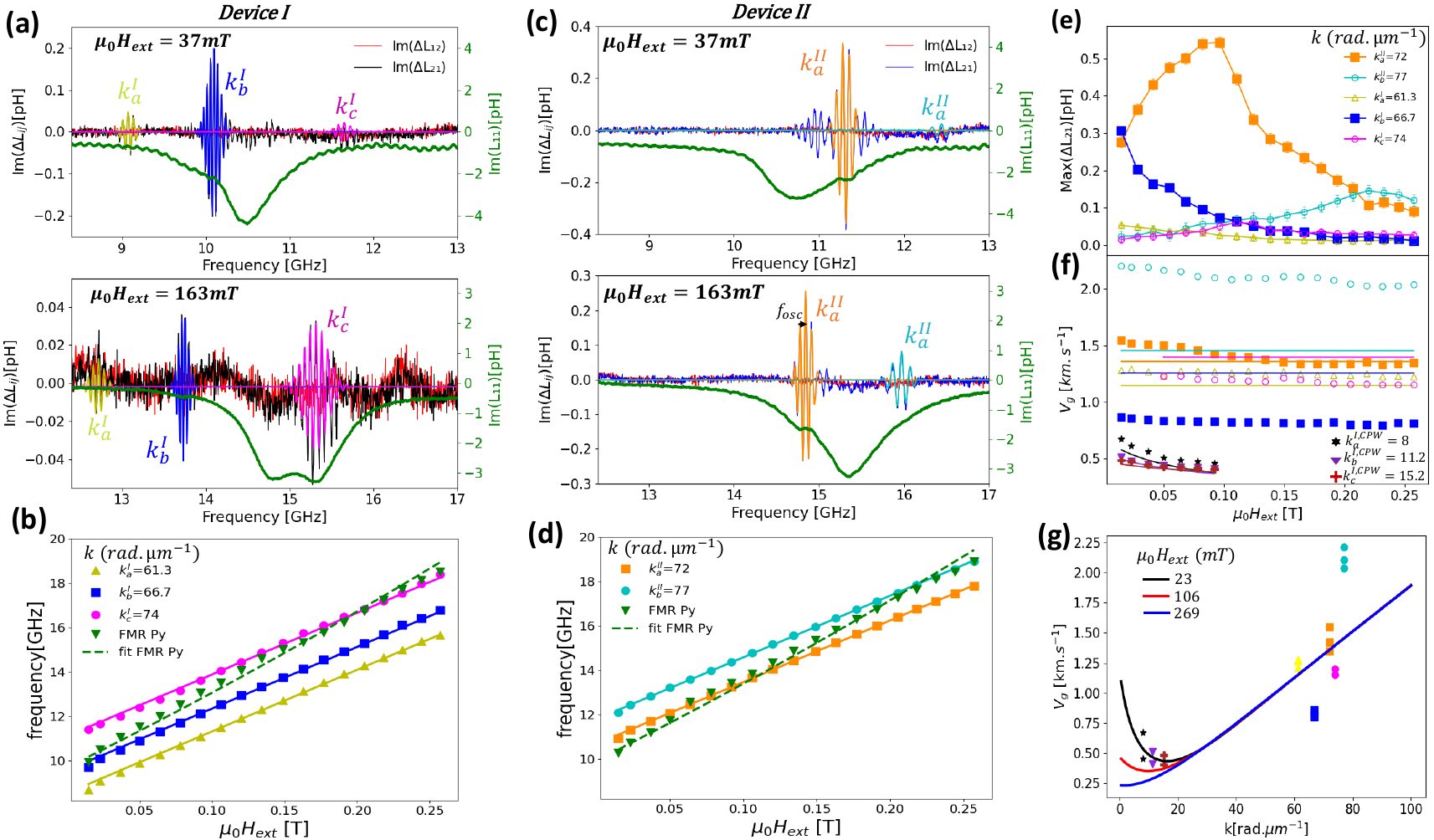}
\caption{\label{fig:epsart} Transmission spectra at 37\,mT and 163\,mT for (a) device\,I and (c) device\,II, blue and red line represent respectively transmission spectra $\Delta L_{21}$ and $\Delta L_{12}$. Field dependence of the frequency for the several mode of (b) device\,I and (d) device\,II. (e) Field dependence of all the mode amplitudes. (f) and (g) Comparison of the measured group velocity with theoretical expression.}
\end{figure*}

\subsection{High-k spin wave beam characterization}

We make use of several distinct peaks to characterize the propagation of high k-vector magnons on both devices I and II. To begin with, we use the Kittel mode to fit the lower branch of the reflection spectra in Fig.2-(a), which corresponds to the quasi k=0 FMR resonance of the YIG happening in the larger portion of the CPW close by the picoprobe and far from the NWA. From this analysis, we extract a value of the gyromagnetic ratio of $\gamma/2\pi$=27.7$\pm0.2$\,GHz.T$^{-1}$ and an effective magnetization $\mu_0$M$_{eff}$=185$\pm1$\,mT, which is identical to the saturation magnetization M$_s$ \cite{Dubs2020}, suggesting no in-plane anisotropy. Then, we track the small reflection peak visible in between the two regions (e.g. the peak at 7.4\,GHz in Fig.2-(d)), which corresponds to the first perpendicular standing spin wave (PSSW) mode, and we fit the difference of the square of the frequencies between the PSSW and the Kittel modes \cite{Kalinikos1986}:
\begin{equation}
\begin{split} 
f_{PSSW}^2-f_{FMR}^2=(\frac{\gamma}{2\pi}\mu_0)^2\,[2M_s \Lambda^2\,\frac{\pi^2}{t^2}\,H_{ext} \\
				+ M_s^2\,\Lambda^2\,\frac{\pi^2}{t^2}(1+ \Lambda^2 \frac{\pi^2}{t^2}) ]
\label{diff_sq}
\end{split} 
\end{equation}
where $\Lambda$=$\sqrt{\frac{2A_{ex}}{\mu_0\,M_s^2}}$ is the exchange length. In doing so, we obtain the following exchange constant A$_{exch}$=3.85$\pm0.1$\,pJ.m\,$^{-1}$. At last, we study the field dependence of the several $k_{NW}$ modes features. For this purpose, we use a simple Gaussian function multiplied by a cosine to fit the oscillations of each transmission peak (see colored fit in Fig.3-(a) and 3-(c)), which gives us the peak position, its amplitude, and the period of oscillation. This leaves us fitting the field dependence of the frequency of each mode with only the wavevector as fitting parameter. \\
We show in the Fig.3 the results of this methodology. For device I with a lattice constant a=500\,nm, we followed three modes which correspond to $k_{a}^{I}$=61.3\,rad.$\mu$m$^{-1}$, $k_{b}^{I}$=66.7\,rad.$\mu$m$^{-1}$, and $k_{c}^{I}$=74\,rad.$\mu$m$^{-1}$; and for device II with a=400\,nm, we tracked two modes $k_{a}^{II}$=72\,rad.$\mu$m$^{-1}$, and $k_{b}^{II}$=77\,rad.$\mu$m$^{-1}$. \\
It may seem curious at first that none of these wavevectors corresponds to an integer value of $\dfrac{2\pi}{a}$. However, the Fourier transform of the NWA dynamic dipolar field distribution, which gives the spectral efficiency of the excitation \cite{Vlaminck2010}, is rather complex to depict as it consists of the modulation of the NWA periodicity with the field distribution of the CPW. We therefore perceive these $k_{NW}$ modes to be neighboring ripples of a convoluted spectral distribution that depends on the lattice periodicity, the width of the nanowire, and the antenna field distribution. \\
We also reported the field dependence of the mode amplitudes in Fig.3-(e), which shows non-monotonous behavior. As one can see in the spectra of Fig.3-(a) and Fig.3-(c), the efficiency of the coupling to a particular mode will be all the stronger that its frequency matches with the one of the NWA FMR. As the gyromagnetic ratio of YIG is smaller than the one of permalloy ($\gamma_{Py}/2\pi$=29.5\,GHz.T$^{-1}$), the NWA dispersion will gradually cross the $k_{NW}$ multimodes dispersion across the field range. Unfortunately, this non-monotonous field dependence of the excitation efficiency makes it arduous to characterize the attenuation of these high k-vector with just one kind of device. A series of devices with different distance D would be more suited for determining the attenuation length of these high-k SWB. \\     
Finally, from the period of oscillation f$_{osc}$ of the transmission spectra, we estimate the group velocity according to v$_{g}$=f$_{osc}$*D \cite{Loayza2018}. We compare our measurement of v$_{g}$ with the theoretical expression in Fig.3-(f) and Fig.3-(g). Although the agreement at lower wavevector is fair, we observe some significant discrepancies in the group velocity among the $k_{NW}$ modes. We foresee that an additional phase delay must occur through the propagation path. Namely, considering that the length of NWA is comparable to the propagation distance D, one can expect the static stray field of the wires to cause some inhomogeneities in the static field distribution between the two antennas.

\section{CONCLUSION}

We carried out a spin wave spectroscopy study on two different 50$\mu$m$^2$ Ni$_{80}$Fe$_{20}$ nanowire arrays resonantly coupled to a continuous 55\,nm thin YIG film. We demonstrated unidirectional transmission of 10\,$\mu$m wide spin wave beams up to 77\,rad.$\mu$m$^{-1}$ in the [8;20]\,GHz frequency range. An attempt to characterize the propagation properties of these high k-vector spin wave beams reveals several peculiarities regarding the modes selection, their coupling efficiency, and possible additional phase lag due to inhomogeneous stray field from the nanowires. These findings serve as a guideline for future miniaturization of nonreciprocal magnonic devices.

\begin{acknowledgments}
The authors acknowledge the financial support from the French National research agency (ANR) under the project \textit{MagFunc}, the Region Bretagne with the \textit{CPER-Hypermag} project, and the Département du Finistère through the project \textit{SOSMAG}. We also want to thanks Bernard Abiven for the implementation of the electromagnet and Guillaume Bourcin for fruitful discussions. CD acknowledges funding by the Deutsche Forschungsgemeinschaft (DFG, German Research Foundation) - 271741898.
\end{acknowledgments}

\section*{Data Availability Statement}

The data that support the findings of this study are available from the corresponding author upon reasonable request.











\end{document}